\documentclass{aastex63}

\newcommand{\Caltech}{\affiliation{TAPIR, Walter Burke Institute for Theoretical Physics, MC 350-17,
    California Institute of Technology, Pasadena, California 91125, USA}}
\newcommand{\Cornell}{\affiliation{Center for Radiophysics and Space
    Research, Cornell University, Ithaca, New York, 14853, USA}}
\newcommand{\WSU}{\affiliation{Department of Physics \& Astronomy,
	Washington State University, Pullman, Washington 99164, USA}}

\newcommand{\AEI}{\affiliation{Max-Planck-Institut fur Gravitationsphysik, Albert-Einstein-Institut, D-14476 Potsdam, Germany}}
\newcommand{\UNH}{\affiliation{Department of Physics \& Astronomy, University of New Hampshire, Durham, New Hampshire 03824, USA}}

\newcommand{\beq}{\begin{equation}}
\newcommand{\eeq}{\end{equation}}
\newcommand{\beqn}{\begin{eqnarray}}
\newcommand{\eeqn}{\end{eqnarray}}

\shorttitle{MC transort in NSNS mergers}
\shortauthors{Foucart et al.}

\begin{document}

\title{Monte-Carlo neutrino transport in neutron star merger simulations}

\correspondingauthor{Francois Foucart}
\email{francois.foucart@unh.edu}

\author{Francois Foucart}\UNH
\author{Matthew D. Duez} \WSU 
\author{Francois Hebert}\Caltech
\author{Lawrence E. Kidder} \Cornell 
\author{Harald P. Pfeiffer} \AEI
\author{Mark A. Scheel} \Caltech

\nocollaboration{6}



\begin{abstract}

Gravitational waves and electromagnetic signals from merging neutron star binaries provide valuable information
about the the properties of dense matter, the formation of heavy elements, and high-energy astrophysics. To fully leverage observations of these
systems, we need numerical simulations that provide reliable predictions for the properties of the matter unbound in these mergers. 
An important limitation of current simulations is the use of approximate methods for neutrino transport that do not converge to a 
solution of the transport equations as numerical resolution increases, and thus have errors that are impossible to quantify.
Here, we report on a first simulation of a binary neutron star merger that uses Monte-Carlo techniques to directly solve the transport 
equations in low-density regions. In high-density regions, we use approximations inspired by implicit Monte-Carlo to greatly reduce the
cost of simulations, while only introducing errors quantifiable through more expensive convergence studies.
We simulate an unequal mass neutron star binary merger up to $5\,{\rm ms}$ past merger, and report on the properties of 
the matter and neutrino outflows. Finally, we compare our results to the output of our best approximate `M1' transport scheme, demonstrating that an
M1 scheme that carefully approximates the neutrino energy spectrum only leads to $\sim 10\%$ uncertainty in the composition and velocity of the ejecta,
and $\sim20\%$ uncertainty in the $\nu_e$ and $\bar\nu_e$ luminosities and energies. The most significant disagreement found between M1 and Monte-Carlo
results is a factor of $\sim 2$ difference in the luminosity of heavy-lepton neutrinos.

\end{abstract}



\section{Introduction} \label{sec:intro}

Gravitational wave (GW) and electromagnetic (EM) observations of neutron star mergers provide us with important information about the properties of dense matter, the synthesis of heavy nuclei, and high-energy astrophysics, as demonstrated by the first detection of GWs from a neutron star merger~\citep{TheLIGOScientific:2017qsa} and associated EM observations (e.g.~\citet{2017ApJ...848L..13A,2017Sci...358.1559K,2017ApJ...848L..19C,2017Natur.551...75S,2017ApJ...848L..16S,Cowperthwaite:2017dyu}). The UV/optical/infrared signal powered by r-process nucleosynthesis in the matter unbound by the merger (kilonova)~\citep{Li:1998bw,Roberts2011} is of particular interest for nuclear astrophysics for the information that it provides about nucleosynthesis in mergers, and about the equation of state of neutron stars. To extract information from observed kilonovae, however, reliable theoretical models are required. These in turn rely on a good understanding of the properties of the matter ejected during and after merger~\citep{2013ApJ...775...18B}, and of nuclear physics in the neutron-rich ejecta~\citep{Barnes:2016}. 

Simulations are our main source of information about merger outflows. However, they suffer from important limitations: they do not capture the growth of magnetic fields from realistic initial strengths~\citep{Kiuchi2015}, and use approximate methods for neutrino transport~\citep{Foucart:2016rxm,Foucart:2018gis}. As a result, they can miss important physical processes: magnetic fields heat the remnant, drive angular moment transport in the system, and produce most post-merger outflows, while neutrinos cool the remnant and drive the evolution of its composition.

For neutrino transport, the main issue is the high dimensionality of the problem. Ideally, one would evolve the neutrino distribution function using Boltzmann's equations of radiation transport. Unfortunately, this is a function of time, position, neutrino energy and momentum, making this a 7-dimensional problem for each neutrino species. The problem is further complicated by the existence of stiff coupling terms between neutrinos and nucleons in dense and hot regions. Merger simulations first included neutrino effects through leakage schemes~\citep{2010CQGra..27k4107S,Deaton2013} that account for the local cooling effects of neutrinos at an order-of-magnitude level. More recently, grey two-moment schemes ('M1' schemes) that evolve the neutrino energy and momentum density but use approximate analytical closures for the neutrino pressure and energy spectrum have been implemented~\citep{shibata:11,Wanajo2014,FoucartM1:2015}, as well as a mixed leakage-one moment scheme~\citep{Radice:2016}. Simulations using M1 schemes have clearly demonstrated that leakage is insufficient to capture the composition of matter outflows in mergers~\citep{Wanajo2014}, the most important parameter to determine the outcome of nucleosynthesis in merger outflows. Yet M1 schemes themselves show that outflow composition and neutrino luminosities have non-negligible dependencies on the exact choice of analytical closures~\citep{Foucart:2016rxm,Foucart:2018gis}, and that standard closures lead to numerical artifacts in simulations (e.g. neutrino shocks in polar regions). Most importantly, while M1 schemes may be sufficient for many purposes, there is no way to test their accuracy without comparison with a solution to the transport equations, as they do not converge to the correct solution when increasing resolution.

In recent years, general relativistic Monte-Carlo (MC) algorithms have risen as a tempting alternative to provide low-cost neutrino transport in merger and post-merger simulations~\citep{richers:15,Ryan2015,Miller:2019dpt}. Building on our implementation of a MC algorithm as a closure to a M1 code~\citep{Foucart:2017mbt}, we present here a first MC transport algorithm for fully general relativistic merger simulations, as well as a first simulation of merging neutron stars with MC transport. The aim of this code is to provide cheap yet reasonably accurate solutions to the transport problem, within a framework that converges to the correct physical solution as more computational resources become available. This code can be used for direct simulations of neutron star mergers, as well as to test the accuracy of existing M1 and leakage results, and can thus greatly improve our understanding of the merger outflows. We note that low-cost MC transport is possible in neutron star mergers because neutrinos impact the evolution of the merger remnant on time scales that are long compared to the simulation time step; e.g. the cooling time scale of a post-merger accretion disk is $\gtrsim (10-100)\,{\rm ms}$~\citep{Deaton2013}. Additionally, neutrino-driven winds have mass outflows of only $\dot M \sim 0.1M_\odot{\rm s^{-1}}$ even immediately after merger~\citep{Foucart:2016rxm}, and remain subdominant with respect to viscous/magnetically-driven winds at later times~\citep{Just2014}. MC sampling errors thus have a limited impact on the dynamics of the remnant.

\section{Methods} \label{sec:methods}

We perform general relativistic radiation hydrodynamics simulations of merging neutron stars with masses $M_1=1.27M_\odot,M_2=1.58M_\odot$, using the 'DD2' equation of state from~\cite{Hempel:2011mk}. The neutron stars have radii $R\sim 13.2\,{\rm km}$ and zero spin. We generate initial data with the {\it Spells} code~\citep{Pfeiffer2003,FoucartEtAl:2008} four orbits before merger, and end the simulations $\sim 5\,{\rm ms}$ after merger. This is sufficient to observe tidal ejection and the production of a neutrino-driven wind. Evolution over longer timescales would require us to model angular momentum transport and heating due to turbulence, through magnetic field evolution or the use of a viscous model. We evolve neutrinos using either our new MC transport scheme or approximate `M1' transport. In this section, we briefly discuss our numerical methods. Readers interested in our results and their implications may skip to Section~\ref{sec:results}.

Simulations are performed with the SpEC code\footnote{https://www.black-holes.org/code/SpEC.html}. SpEC evolves Einstein's equations in the Generalized Harmonics formalism~\citep{Lindblom2006} using pseudospectral methods with adaptive mesh refinement~\citep{Szilagyi:2014fna}. The fluid equations are evolved on a Cartesian grid, using high-order finite volume shock-capturing methods. We use the methods of~\cite{Duez:2008rb,Foucart:2013a}, except that we allow the time step on the pseudospectral grid to be smaller than the time step on the finite volume grid. At the resolution used in this manuscript, time stepping errors remain small compared to other sources of errors. The finite volume grid has a spacing $\Delta x_{\rm FV}=188\,{\rm m}$ at the beginning of the evolution, and $\Delta x_{\rm FV}=200\,{\rm m}$ after merger. We use fixed mesh refinement after merger, doubling the grid spacing at each level. Each of our 4 refinement levels has $200\times 200 \times 176$ points. 

We perform three simulations. The first uses the two-moment ('M1') transport scheme from~\cite{FoucartM1:2015,Foucart:2016rxm}, evolving the neutrino energy density, momentum density, and number density. It uses approximate analytical closures to estimate the pressure tensor and energy spectrum. The others use MC transport, with different numbers of packets in order to estimate sampling errors in the simulations. The core of our MC algorithm is described in~\cite{Foucart:2017mbt}. Here, we summarize the main components of the algorithm, and new features needed to obtain stable and accurate evolution in merging neutron stars. 

All simulations assume that neutrino-matter interactions can be described by an emissivity $\eta$, absorption opacity $\kappa_a$, and elastic scattering opacity $\kappa_s$. We use tabulated values produced with the NuLib library~\citep{OConnor:2015}. The table includes reaction rates for the charged current reactions 
\beq
p + e^- \leftrightarrow n+\nu_e \hspace{1cm} n+e^+ \leftrightarrow p+\bar\nu_e;
\eeq
scattering of neutrinos on protons, neutrons, $\alpha$-particles and heavy nuclei; and, for the muon and tau (anti)neutrinos only, $e^+e^- \leftrightarrow \nu\bar \nu$ and Bremsstrahlung. Under these assumptions, muon and tau (anti)neutrinos all behave in the same manner, and we treat them as a single species (called $\nu_x$). The table is logarithmically spaced in neutrino energies (16 groups up to $E=528\,{\rm MeV}$), density (86 points in $[1e6,3.2e15]\,{\rm g/cm^3}$) and temperature (65 points in $[0.05,150]\,{\rm MeV}$), and linearly spaced in the electron fraction $Y_e$ (51 points in $[0.01,0.6]$). 

The basic idea of the MC method is to sample, for each neutrinos species, the distribution function $f_{(\nu)}(t,x^i,p^\mu)$ using a discrete number $P$ of packets that each represent a number $N$ of neutrinos. More precisely, 
\beq
f_{(\nu)}(t,x^i,p_\mu) \approx \sum_{k=1}^P N_k \delta^3(x^i-x^i_k) \delta^3(p_i-p_i^k) 
\eeq
with $(t,x^i)$ the coordinate time and position, $p^\mu$ the 4-momentum, $(x^i_k,p_i^k)$ the position and spatial components of the momentum one-form of packet $k$ at time $t$, and $N_k$ the number of neutrinos that this packet represents. Packets are created from an isotropic distribution in the fluid frame, propagated along null geodesics, and scattered/absorbed with probabilities set by $\kappa_a,\kappa_s$ (see ~\cite{Foucart:2017mbt}). We use a split operator method where the fluid and metric are evolved first, and neutrino packets second. We also allow the MC code to take time steps covering multiple fluid steps when possible: the MC code aims to take steps with $c\Delta t = (0.5-1)\Delta x_{\rm FV}$. Neutrinos deposit/remove energy and momentum from the fluid at the end of the MC step. For the interactions considered here, changes in the fluid variables due to neutrino-matter interactions follow the equations for conservation of energy, momentum, and lepton number:
\beq
\nabla_\mu T^{\mu\nu}_{\rm fl} = -\eta u^\nu +\sum_k \left(\kappa_a J_k u^\mu + [\kappa_a+\kappa_s] H_k^\mu\right);\,\,\nabla_\mu \left(\rho_0 Y_e u^\mu\right)=-\sum_s s_s\eta_{N,s} + \sum_k s_k \kappa_a \frac{J_k}{\nu_k} 
\label{eq:coupling}
\eeq
where $T^{\mu\nu}_{\rm fl}$ is the stress-energy tensor of the fluid, $\eta$ the total neutrino energy emissivity, $\eta_{N,s}$ the number emissivity of neutrinos of species $s$, $u^\mu$ the fluid 4-velocity, $J_k,H^\mu_k$ the energy density and momentum density of neutrinos in packet $k$, and $\nu_k$ their average energy, in the fluid frame. The electron fraction $Y_e=n_p/(n_p+n_n)$ (with $n_{p,n}$ the number density of protons and neutrons) parametrizes the composition of the fluid. $s=1$ for electron neutrinos, $-1$ for electron antineutrinos, and $0$ otherwise. We discuss below how these terms are estimated. 

The main components of this code were used in~\cite{Foucart:2018gis} to estimate errors in the M1 method. However, at the time we could not use MC methods in the densest, hottest regions of the merger. Hot regions are problematic for MC algorithms as large emissivities and opacities cause rapid creation and destruction of packets, and numerical instabilities. To avoid this, we adapt to the merger problems the ideas of implicit MC, partially following the work of~\cite{1971JCoPh...8..313F}. Once the absorption opacity becomes too large, the emissivities and absorption coefficients are modified according to
\beq
\kappa_a' = (1-\alpha) \kappa_a;\,\, \kappa_s' = \kappa_s + \alpha \kappa_a ;\,\, \eta' = (1-\alpha)\eta
\eeq
for some constant $\alpha$, thus reducing $(\eta,\kappa_a)$ without modifying the neutrino diffusion rate or equilibrium energy density. As opposed to~\cite{1971JCoPh...8..313F}, we choose $\alpha$ separately for each species and energy bin. We require that $\kappa_a' \Delta t<0.5$, effectively guaranteeing that the equilibration time scale is always at least a few time steps. We also require
\beq
\alpha > \frac{\beta \Delta t \kappa_a}{1+\beta \Delta t \kappa_a};\,\, \beta=\rm{min}\left(\frac{du_{\rm rad}}{du_{\rm fl}}|_{(\rho,Y_e)},\frac{dn_{rad}}{dn_{\rm fl}}|_{(\rho,T)} \right)
\eeq
with $u_{\rm fl,rad}$ and $n_{\rm fl,rad}$ the energy density and lepton number density of the fluid and of neutrinos in equilibrium with that fluid. This aims to prevent instabilities in the joint evolution of the fluid and neutrino radiation. While implicit MC in~\cite{1971JCoPh...8..313F} was specifically designed to match a given time discretization of the original transport equations, our scheme only does so in the limit $\Delta t\rightarrow 0$. This is because we couple neutrinos to both the composition and internal energy of the fluid (as opposed to the energy only for photons), and because~\cite{1971JCoPh...8..313F} used the same $\alpha$ for all energy groups, and an effective {\it inelastic} scattering cross-section such that neutrinos have an equilibrium energy spectrum post-scattering. 

As we fix the number of MC packets emitted per unit time (see below), the effective number of packets in optically thick cells is $\propto (\kappa_a')^{-1}$. Decreasing $\kappa_a'$ allows us to reduce statistical errors without increasing the cost of simulations, while introducing a small error that converges away with resolution. To test this method in circumstances reasonably similar to neutron star merger conditions, we consider the evolution of a post-bounce supernova remnant, performed with much coarser
resolution than our merger simulations~\citep{Abdikamalov2012,Foucart:2017mbt}. We find that the $\nu_e$ and $\bar \nu_e$ luminosities are very accurate, while the $\nu_x$ luminosity is impacted at the $\sim 20\%$ level. This gives us confidence that even in an unfavorable configuration for this approximation, the method provides reasonably accurate results. Nevertheless, improving this part of the algorithm without drastically increasing the cost of simulations or causing instabilities may be useful in the future. We note that transforming the absorption opacity into a scattering opacity would not provide much of a gain if we always treated scattering events explicitly. However, we developed in~\cite{Foucart:2017mbt} a cost-effective method to approximate many scatterings as a diffusion process. A different implementation of that idea is also used by~\cite{richers:15}.

To increase stability, we also compute $\eta,\kappa_a,\kappa_s$ using {\it predicted values} for the temperature and composition of the fluid. These are computed by solving implicitly the energy and lepton number conservation equations (Eq.\ref{eq:coupling}), using the expectation values of the source terms and neglecting changes in the momentum of the fluid. 

Coupling terms between neutrinos and the fluid are evaluated by integrating the right-hand sides of Eq.~\ref{eq:coupling} over a MC time step. This requires integrals, for each packet, of $J_k$ and $H^\mu_k$ over the worldline of a packet. In the MC formalism, and for any time interval $\Delta t$ between neutrino-matter interactions, these are simply (see~\cite{Foucart:2017mbt})
\beq
\int dt J_k =  \Delta t N_k \frac{\nu_k^2}{p^t_k \sqrt{-g}\Delta V};\,\, \int dt  H^\mu_k =\Delta t N_k \frac{\nu_k}{p^t_k \Delta V\sqrt{-g}} p_\nu^k (g^{\mu\nu} + u^\mu u^\nu),
\eeq
with $\delta^3(x^i_k-x^i)= (\Delta V)^{-1}$ in the finite volume discretization, and $\Delta V$ the coordinate volume of a cell.  

\begin{figure}
\centering
\includegraphics[width=0.8\columnwidth]{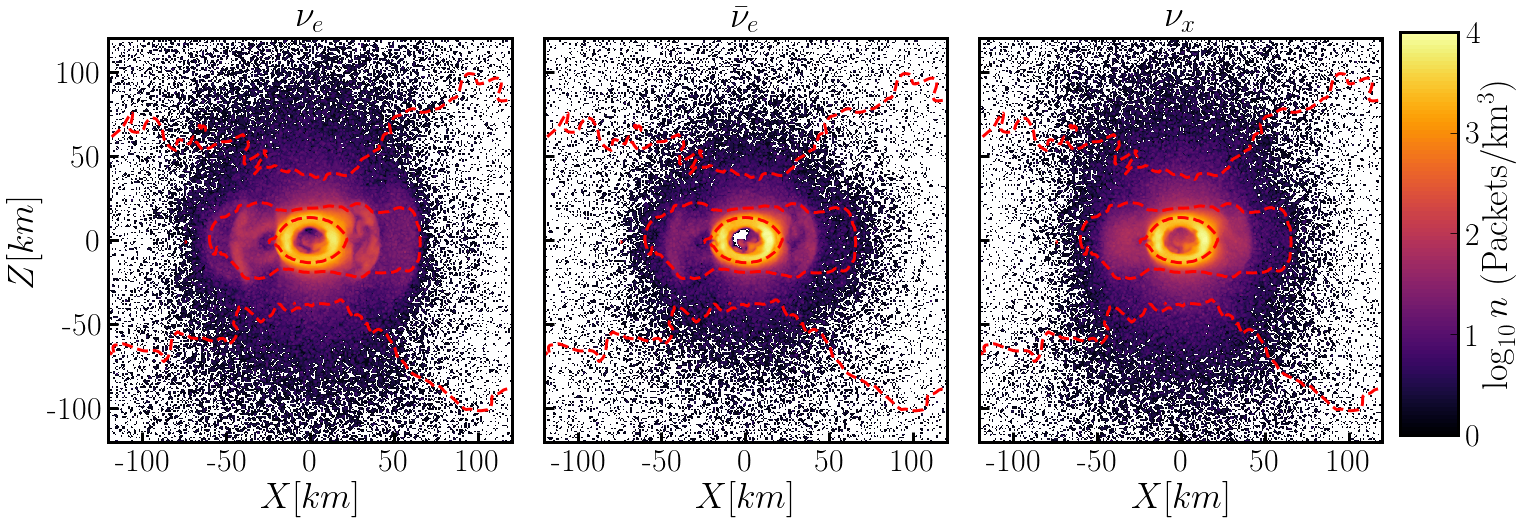}
  \caption{Number density of MC packets for each species, in a vertical slice at the end of the high-resolution simulation. The dashed red lines are density contours at $10^{9,11,13}\,{\rm g/cm^3}$. Fig.~\ref{fig:Ye} shows the same snapshot. Most packets are in the hot region close to the stellar surface. Few packets are needed far away from the remnant, or in cold regions of the neutron star.}
\label{fig:N}
\end{figure}

Finally, we need to choose the number of neutrinos $N_k$ represented by a packet. We could take the total energy $E_k=N_k\nu_k$ of a packet to be constant, or fix the total number of packets. However, none of these choices proved sufficient to limit noise in our low-cost simulations. Instead, we choose a {\it minimum energy} ($E_{\rm min}$), a {\it desired number of packets to be emitted} within a grid cell per light-crossing time of the cell ($N_{\rm em,target}$), and a {\it maximum number of packets} per species over the whole domain ($N_{\rm max,tot}$). We first set $E_k$ to get the desired $N_{\rm em,target}$, and reset it to $E_{\rm min}$ if $E_k<E_{\rm min}$. Whenever the number of packets of a given species grows above $N_{\rm max,tot}$, we increase $E_{\rm min}$ for that species by $10\%$, and resample the existing packets: $10\%$ of the packets are randomly destroyed, while the surviving packets multiply their $N_k$ by $(1.1)^{-1}$. This guarantees a minimum number of packets per cell in hot regions, without reducing too much the number of packets elsewhere. We use $N_{\rm max,tot}=(12,3)\times 10^7$, $E_{\rm min}=(1,4)\times 10^{-14}M_\odot c^2$, and $N_{\rm em,target}=(100,25)$ for our two MC simulations. The factor of $4$ between simulations halves the expected sampling noise.
Even with that method, a majority of packets are concentrated in the hottest regions (see Fig.~\ref{fig:N}). Accordingly, instead of evolving MC packets on the processor responsible for the fluid cell in which they are located (as in~\cite{Foucart:2018gis}), our code can offload the evolution of all MC packets within a given finite volume cell to another processor, as needed for load-balancing.

\section{Simulation Results} \label{sec:results}

\begin{figure}
\centering
\includegraphics[width=\columnwidth]{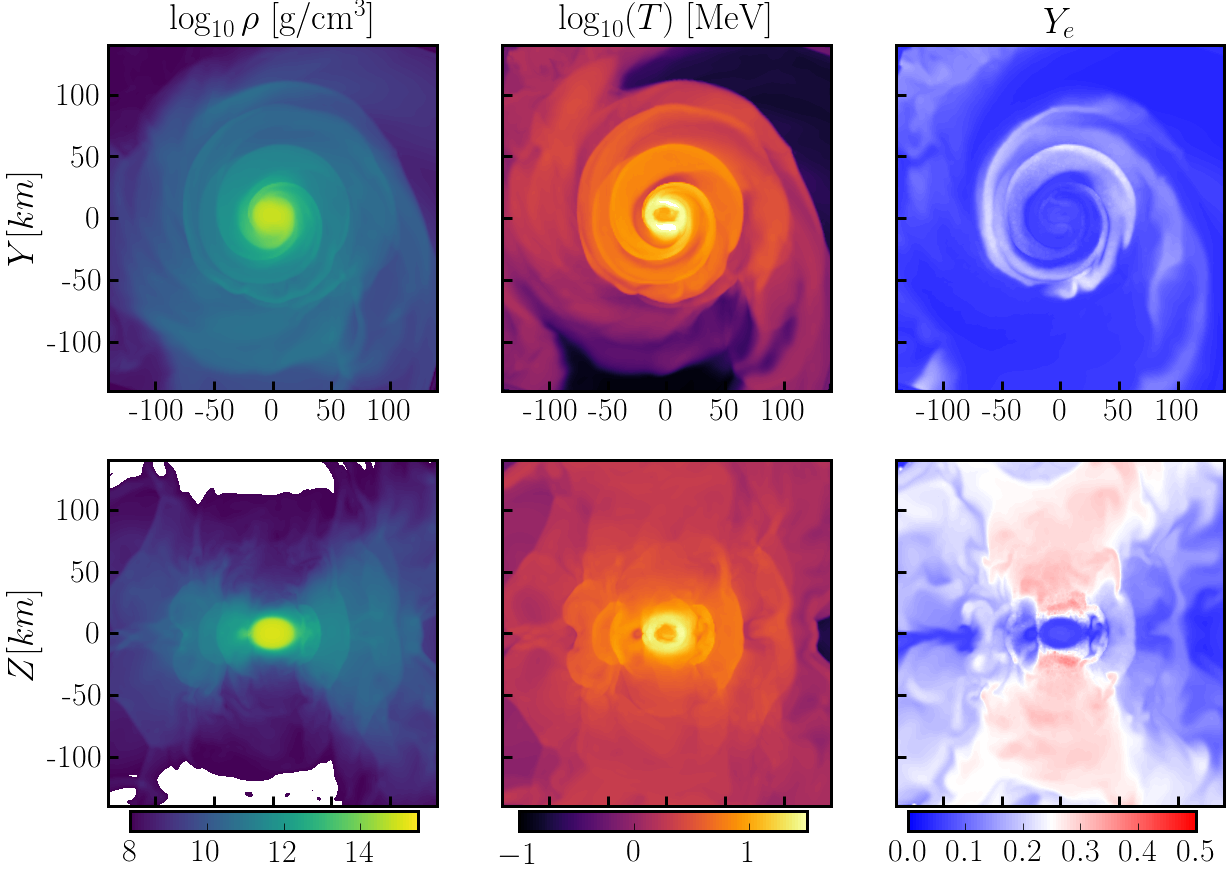}
  \caption{Snapshot of the MC simulation with the largest number of packets, $5\,{\rm ms}$ after merger. We show the density ({\it Left}), temperature ({\it Center}), and electron fraction ({\it Right}) in the equatorial plane ({\it Top}) and
  a vertical slice passing through the center of the remnant ({\it Bottom}). We can see the post-merger hypermassive neutron star at the center of the figures, surrounded by a dense torus. Low-density outflows are launched mostly along the edges of the torus.}
\label{fig:T5}
\end{figure}

The qualitative features of the evolution are similar in all simulations. The neutron stars perform $\sim 4$ orbits before merging. Due to the mass asymmetry, a significant amount of mass is unbound in a cold, neutron-rich tidal tail, while hotter, less neutron-rich matter is ejected following the collision of the neutron star cores. Within the following $\sim 5\,{\rm ms}$, a massive torus forms around the dense merger remnant, while a neutrino-driven wind develops along the edges of the torus and in the polar regions. Figure~\ref{fig:T5} shows the main properties of the remnant $5\,{\rm ms}$ after merger: the density, temperature, and electron fraction. The compact remnant reaches temperatures up to $\sim 60\,{\rm MeV}$, and remains neutron rich ($Y_e\lesssim 0.1$). The surrounding torus has temperature of a few MeVs and higher $Y_e$ (0.15-0.25), while the polar outflows are even less neutron rich. The main roles of neutrinos are to cool the system and, for $\nu_e$ and $\bar \nu_e$, drive changes in $Y_e$. Over longer time scales, we expect the remnant to form a uniformly rotating star surrounded by a more axisymmetric disk. A large fraction of that disk ($\gtrsim 30\%$) is unbound by magnetically-driven winds, viscous angular momentum transport, and neutrino-driven winds, with neutrinos playing a major role in setting $Y_e$ in the outflows~\citep{Metzger2014,Fujibayashi:2020dvr}.

\begin{table}
\caption{
  Matter outflows in our 3 simulations. We provide the total ejected mass, average $Y_e$ of the ejecta, and average asymptotic velocity of the ejecta. We also provide these same quantities for the polar ejecta, defined as unbound material with a velocity vector inclined by less than $45^\circ$ with respect to the rotation axis.
}
\label{tab:outflows}
\begin{tabular}{c|ccc|ccc|}
{\rm Sim} & $M_{\rm ej}\,(10^{-3}M_\odot)$  & $\langle Y_e\rangle_{\rm ej}$ & $\langle v\rangle_{\rm ej}$ & $M_{\rm ej,pol}\,(10^{-3}M_\odot)$ &  $\langle Y_e\rangle_{\rm ej,pol}$ & $\langle v\rangle_{\rm ej,pol}$\\
\hline
MC-low & 11.56 & 0.135 &0.214c & $0.53$ & 0.228 & 0.192c \\
MC-high & 8.25 & 0.130 & 0.206c & $0.49$ & 0.234 & 0.191c\\
M1 &  8.31 & 0.129 &0.201c & $0.34$ & 0.259 & 0.184c  \\
\end{tabular}
\end{table}

\begin{figure}
\centering
\includegraphics[width=\columnwidth]{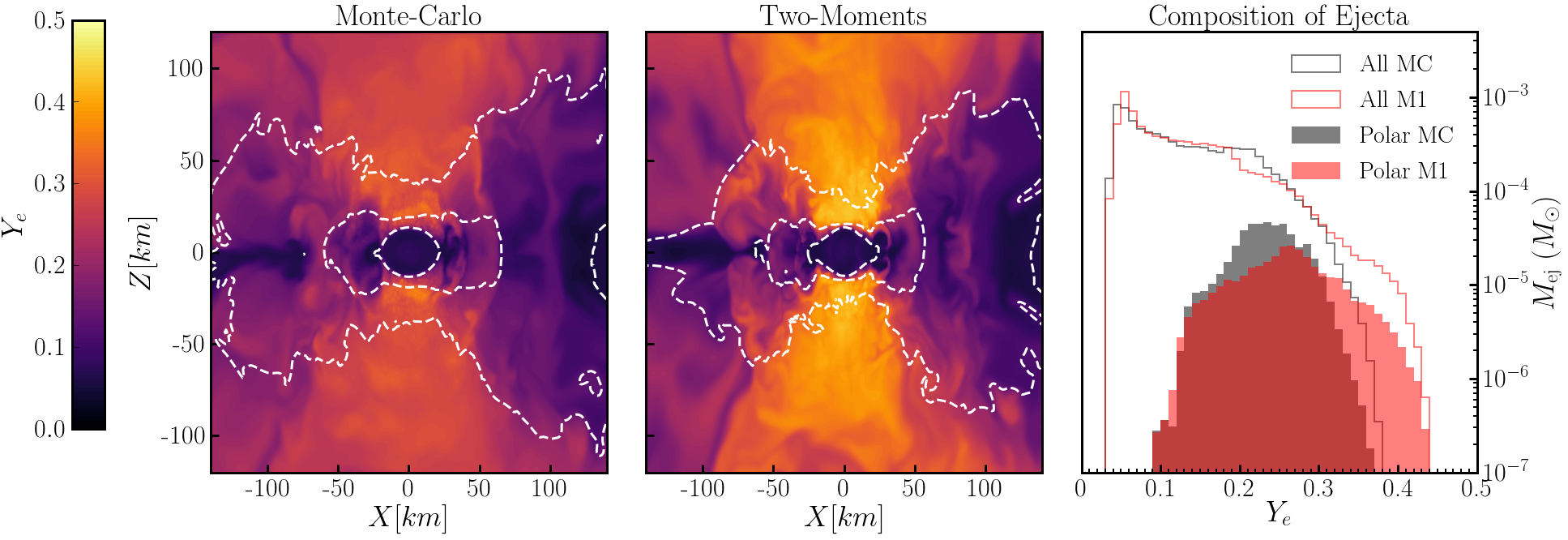}
  \caption{{\it Left}: Electron fraction of the remnant $5\,{\rm ms}$ after merger in the MC simulation with the highest packet count. Dashed white lines are density contours at $10^{9,11,13}\,{\rm g/cm^3}$ {\it Center}: The same, for the M1 simulation. Regions just outside of the neutron star and disk are slightly denser in the MC simulations (possibly due to he higher $\nu_x$ luminosity), and matter is more neutron rich in the outflow regions. {\it Right}: Composition of the unbound material. Filled histograms shows the polar outflows only.}
  \label{fig:Ye}
\end{figure}

\begin{figure}
\centering
\includegraphics[width=0.7\columnwidth]{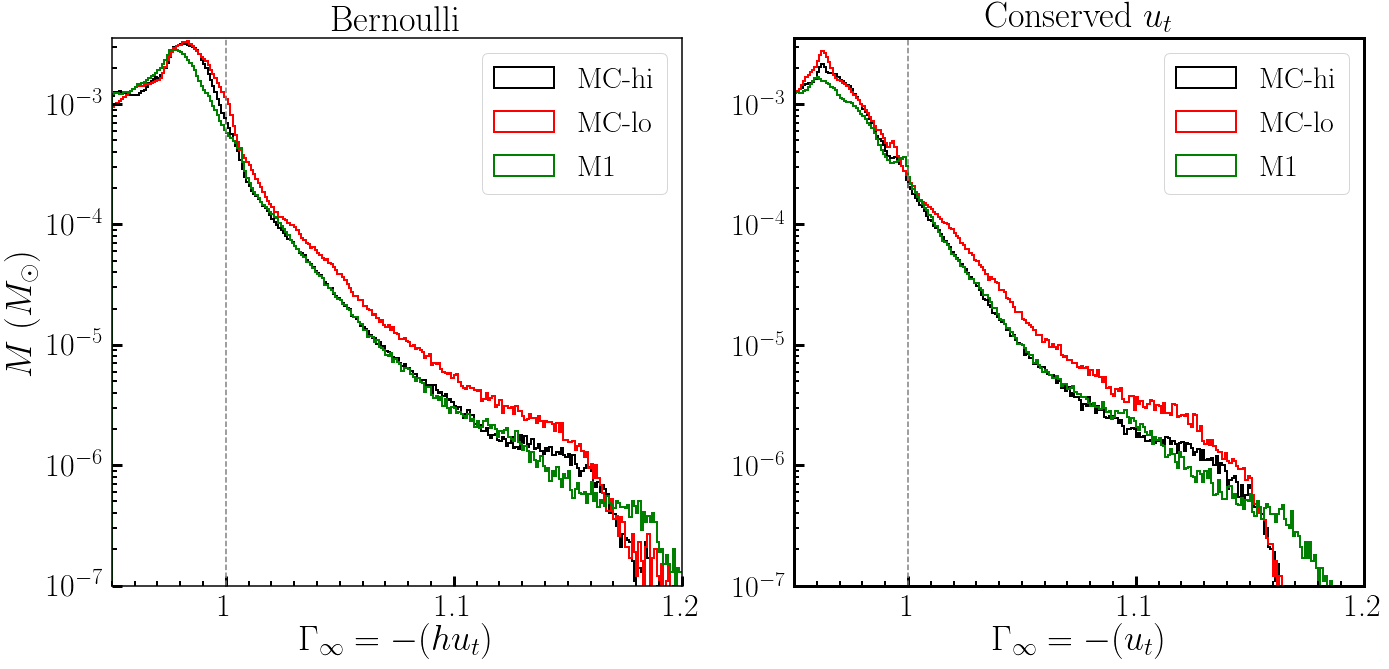}
  \caption{Estimated Lorentz factor of the ejecta at infinity. We show results using the Bernoulli criteria ({\it Left}), which slightly overestimates ejected masses, and assuming that $u_t$ is constant ({\it Right}), which typically underestimates the ejected mass. $\Gamma_\infty=1$ (dashed grey line) is the estimated boundary between bound and unbound material.}
  \label{fig:W}
\end{figure}

The mass, composition, velocity, and geometry of the outflows are of particular interest, as they set the main observable properties of kilonovae~\citep{2013ApJ...775...18B}. High-$Y_e$ outflows produce optical kilonovae evolving on timescales of days and do not produce the heaviest r-process nuclei, while low-$Y_e$ outflows produce infrared, week-long kilonovae that mostly produce heavier nuclei. The (approximate) boundary between these two outcomes is $Y_e\sim 0.25$~\citep{Lippuner2015}. The mass and velocity of the outflows also impact the brightness and duration of kilonovae. 
Figs.~\ref{fig:Ye}-\ref{fig:W} and Table~\ref{tab:outflows} summarize the main properties of the outflows. While there are some meaningful differences between simulations, these are generally smaller than when comparing our 'best' M1 scheme (used here) to a simpler version with an approximate treatment of neutrino energies~\citep{Foucart:2016rxm}, and dwarfed by differences between neutrino transport and leakage schemes~\citep{Wanajo2014,Foucart:2015gaa}. This is an encouraging sign for both M1 and MC schemes. The main differences that cannot be explained by MC sampling errors are a $\sim 10\%$ lower value of $Y_e$ in the polar outflows ($<45^\circ$ from the rotation axis), and a lower cutoff for the maximum value of $Y_e$ reached by the outflows.
This difference grows in a $\sim 5\,{\rm km}$ region along the polar axis, around and outside of the neutrinospheres of $(10-20)\,{\rm MeV}$ electron (anti)neutrinos. In that region, $Y_e$ is mostly constant in the MC simulation, but increases in the M1 simulation. One possible explanation is that this region is farther from the remnant in the MC simulation (Fig.~\ref{fig:Ye}), and that the M1 closure increases the density of neutrinos at the poles -- so that even though the neutrino luminosities are higher in the MC simulation, neutrino fluxes are smaller. However, as this is a region with rapid variations of the fluid properties, significant neutrino pressure, and non-negligible impacts of the M1 energy closure, other potential sources of errors cannot be ignored.

The total mass of unbound material (Table~\ref{tab:outflows}) and estimated velocity distribution of the outflows (Fig.~\ref{fig:W}) show broad agreement between simulations, with just two noticeable differences. First, the MC simulation with a lower number of packets overproduce outflows by $\sim 30\%$. This is less than e.g. the factor of $2$ uncertainty associated with the definition of the ejecta itself (see Fig.~\ref{fig:W}), but it is the most noticeable effect of MC sampling errors. The excess ejecta has $Y_e\sim 0.2$, is close to the orbital plane ($\theta>60^\circ$), and is observed $(1-2)\,{\rm ms}$ post-merger and in the final snapshot. It is thus most likely ejected during core bounces, from the hottest regions of the remnant, where neutrino pressure is the most significant. This would most naturally explain the properties of the excess ejecta, and the fact that the neutrino luminosity surprisingly varies less with MC sampling rate than the ejected mass (see below).
Fig.~\ref{fig:W} also shows why it is so difficult to accurately estimate the amount of unbound material in a simulation: the steepness of the Lorentz factor distributions implies that a very small error in the estimated location of the boundary between bound and unbound material has large effects on the predicted mass of unbound material. The second, more minor difference is that the cutoff of the velocity distribution is slightly lower in both MC simulations ($v_{\rm max}\sim 0.5$) than with M1 ($v_{\rm max}\sim 0.55$).
We can compare these results to~\cite{Sekiguchi:2016}, who performed a longer simulation with M1 transport, the same equation of state, and slightly more symmetric masses. They find $0.005M_\odot$ of ejecta with $\langle Y_e \rangle = 0.2$ and $\langle v \rangle = 0.19$, and a $Y_e$ distribution fairly similar to those obtained with our M1 code  -- results that are broadly consistent if the higher mass asymmetry led to the ejection of an additional $0.003M_\odot$ of neutron-rich material.

\begin{figure}
\centering
\includegraphics[width=\columnwidth]{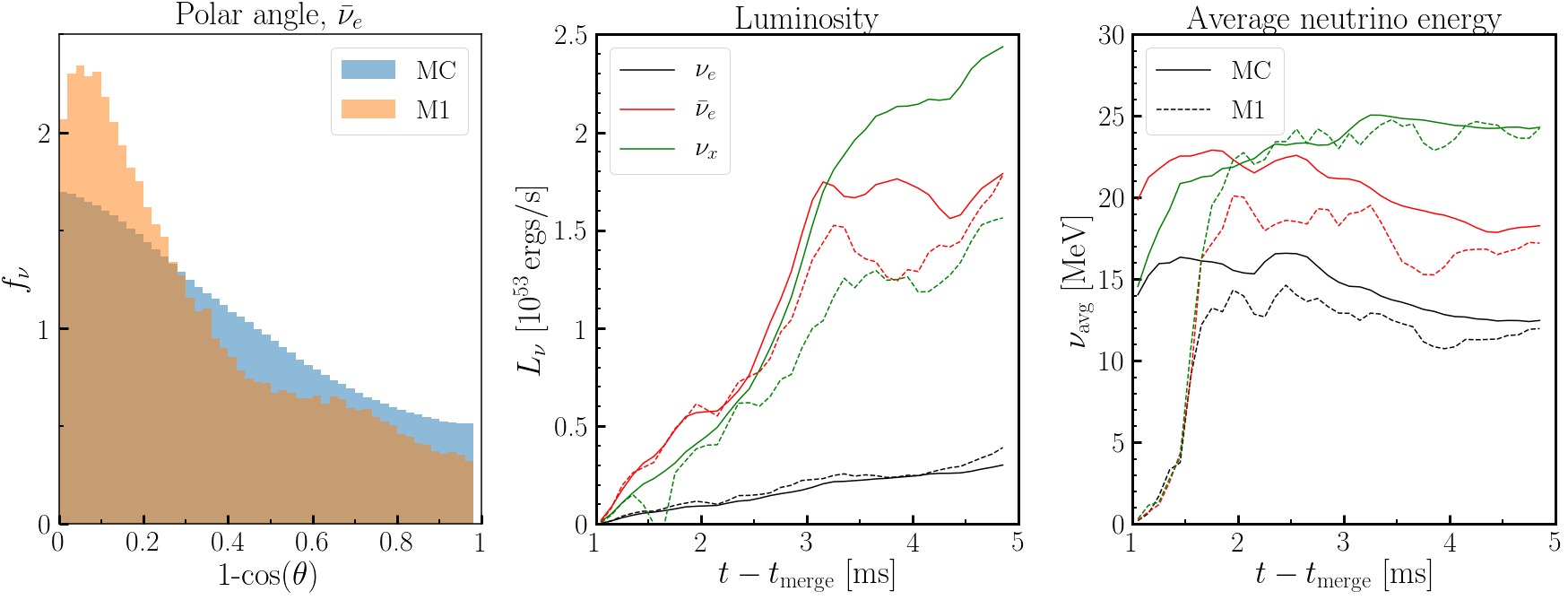}
  \caption{{\it Left}: Angular distribution of the electron antineutrinos leaving the grid $(3-5)\,{\rm ms}$ after merger; $\theta$ is the angle with respect to the angular momentum vector, i.e. polar neutrinos are on the left, equatorial neutrinos on the right, and each bin covers the same surface area. {\it Center}: Neutrino luminosity in the high resolution MC simulation and the M1 simulation. {\it Right}: Average (number-weighted) energy of neutrinos leaving the computational grid in the same simulations.}
  \label{fig:NuProp}
\end{figure}

The properties of neutrinos escaping the system are summarized in Fig.~\ref{fig:NuProp}. After the first $2\,{\rm ms}$, the average energy of neutrinos and the $\nu_e$ and $\bar\nu_e$ luminosities in the M1 and MC simulations agree within $\sim 20\%$. The $\nu_x$ luminosity, on the other hand, is nearly twice as large in the MC simulation. Differences between the two MC simulations are negligible compared to differences between MC and M1. The luminosity of heavy-lepton neutrinos is one of the most uncertain observable in our M1 scheme, because a large region close to the neutron star surface has negligible $\kappa_a$ but large $\kappa_s$. In that region, neutrinos are trapped, but out of thermal equilibrium with the fluid. As the energy closure is the most ad-hoc part of our M1 algorithm, and the diffusion rate of neutrinos strongly depends on the choice of energy spectrum, this is a particularly difficult situation. The MC scheme has no particular reason to perform poorly in that regime: while it corrects large absorption opacities, its treatment of high-scattering regions is better motivated than that of the grey M1 scheme. Nevertheless, it may be premature to assume that the MC scheme provides the better answer, as the two schemes may be impacted in different ways by the grid resolution. In~\cite{Sekiguchi:2016}, neutrino luminosities reach values similar to those measured here within $\sim 5\,{\rm ms}$, before decreasing on the cooling time scale of the remnant.

The angular distribution of neutrinos shows clearer differences between simulations (Fig.~\ref{fig:NuProp}), and is much better captured by the MC algorithm than by the M1 algorithm. In polar regions, there is a nearly $50\%$ excess of neutrinos in the M1 scheme, due to artificial radiation shocks caused by the analytical closure~\citep{Foucart:2018gis}. This mildly impacts the composition of the winds, and would lead to large errors in the calculation of the energy deposited by $\nu\bar\nu$ annihilation in polar regions.

Finally, we report on simulation costs: the two MC simulations cost $230k$ and $310k$ CPU-hrs on the Frontera cluster, at $30k$ and $40k$ CPU-hrs per millisecond at the end of the evolution (the merger phase is costlier). The M1 simulation is not directly comparable at early times, as it was performed on the Comet cluster, but it evolved at $35k$ CPU-hrs per millisecond by the end of the evolution on Frontera. The new MC code is thus competitive with our best M1 code, and cheap enough to be used for at least small parameter space surveys of neutron star mergers on $(10-20)\,{\rm ms}$ time scales, or for a small number of longer evolutions. This is because a small number of MC packets is sufficient to capture the most important neutrino effects: our simulations use only $\sim (1,4)$ packets per finite volume cell.

\section{Implications for numerical relativity and astrophysics}

The first general relativistic simulations of binary neutron stars with MC radiation transport reported here demonstrates that cheap, yet reasonably accurate evolution of the transport equations with MC methods is possible. The only noticeable effect of MC sampling noise in the most important observables of our simulations is a $\sim 30\%$ increase in the outflow mass in our lower accuracy MC simulation. This is comparable to other sources of error in current simulations, even for the very low number of neutrino packets used in this study. The composition and velocity of the outflows, and the neutrino luminosities and energies are largely unaffected by sampling noise. For the specific binary studied in this manuscript, we find $\sim 0.008M_\odot$ of neutron-rich equatorial ejecta, mostly in a tidal tail, as well as an incipient neutrino-driven wind with higher electron fraction $\langle Y_e\rangle\sim 0.23$. Both ejecta components have average velocity $\sim 0.2c$. 
Detailed information about the properties of the ejected nuclear matter and the neutrino outflows are available upon request.
Our MC simulations also provide full snapshots of the neutrino and matter distribution every $0.5\,{\rm ms}$ that can be used for studies of neutrino physics in mergers.

A first comparison of MC results with our approximate M1 code provides generally reassuring results for both methods, with agreement at the $\sim 20\%$ level for the neutrino luminosities (except $\nu_x$) and energies, and very good agreement in the outflow masses and velocity between the M1 simulation and our most accurate MC simulation. The M1 simulation overestimates the average and maximum electron fraction of the outflows by $\sim 10\%$, and the two methods disagree by a factor of $2$ on the luminosity of heavy-lepton neutrinos, but other observables are in remarkable agreement given the resolution of the simulation and the number of MC packets used. This indicates that earlier studies comparing our latest M1 code with more approximate M1~\citep{Foucart:2016rxm} and leakage methods~\citep{Foucart:2015gaa}, that found more significant differences, provide good estimates of the uncertainties of simpler neutrino treatments. Differences in $\nu_x$ luminosities may however have more of an impact if $\nu_x$ oscillate into $\nu_e,\bar \nu_e$ (e.g. due to neutrino-matter resonances [\cite{Vlasenko:2018irq}]), and thus affect $Y_e$. Our new MC algorithm has a low computational cost, will allow us to add new neutrino physics to simulations (e.g. pair annihilation, inelastic scatterings,...), and converges to a solution of the transport equations. On the other hand, the fact that an advanced M1 code is shown to already provide reasonably accurate results is reassuring for current outflow and kilonova models.

\acknowledgments

We are grateful to Sherwood Richers, Ernazar Abdikamalov, Ben Ryan, Charles Gammie, Dan Kasen, Jonah Miller and Josh Dolence for useful discussions regarding neutrino transport and Monte-Carlo methods. We also thank Aurore Bertranhandy for her help with the NuLib library. We are grateful to the Yukawa Institute for Theoretical Physics the Center for Computational Astrophysics for hosting some of these discussions. 
FF gratefully acknowledges support from the DOE through Early Career award DE-SC0020435, from the NSF through grant PHY-1806278, and from NASA through grant 80NSSC18K0565. M.D gratefully acknowledges
support from the NSF through grant PHY-1806207.
F.H. and M.S. acknowledge funding from NSF Grants PHY–1708212 and PHY–1708213. L.K. acknowledges support from NSF grant
PHY-1606654 and PHY-1912081. F.H., L.K. and M.S. also thank the Sherman Fairchild Foundation for their support.

\bibliography{References/References}{}
\bibliographystyle{aasjournal}



\end{document}